\documentclass[aps,twocolumn,showpacs]{revtex4}
\usepackage{amssymb}
\usepackage{mathrsfs}
\usepackage{latexsym}
\usepackage{natbib}
\usepackage{dsfont}

\newcommand{\nn}{\nonumber}

\newcommand{\tr}{\text{Tr}}

\def\xx{\mathbf{x}}

\def\x{x}

\begin{document}

\title{Retarded resonance Casimir-Polder interaction of a uniformly rotating two-atom system  }

\author{Saptarshi Saha}
\email{ss17rs021@iiserkol.ac.in}

\author{Chiranjeeb Singha}
\email{cs12ip026@iiserkol.ac.in}

\author{Arpan Chatterjee}
\email{ac17rs016@iiserkol.ac.in}

\affiliation{ Department of Physical Sciences, 
Indian Institute of Science Education and Research Kolkata, Mohanpur - 741 246, WB, India }
 


\begin{abstract}

We consider here, a two-atom system is uniformly moving
through a circular ring at an ultra-relativistic speed and weakly
interacting with common external fields. The vacuum fluctuations of the
quantum fields generate the entanglement between the atoms. Hence an
effective energy shift is originated, which depends on the inter-atomic
distance. This is commonly known as resonance Casimir-Polder
interaction (RCPI). It is well known that, for a linearly accelerated
system coupled with a massless scalar field, we get a thermal response
when the local inertial approximation is valid. On the contrary, the
non-thermality arises in the presence of the centripetal acceleration.
We use the quantum master equation formalism to calculate the
second-order energy shift of the entangled states in the presence of
two kinds of fields. They are the massive free scalar field and the
electromagnetic vector field. For both cases, we observe the
non-thermal behavior. A unique retarded response is also noticed in
comparison to the free massless case, which can be observed via the
polarization transfer technique.

\end{abstract}

\maketitle

\section{Introduction} \label{introduction}

A remarkable distinction of classical to the quantum domain is the existence of the zero-point energy
\cite{Wheeler1945}. Feynman pointed out that the change in the zero-point
energies give rise to the Lamb shift in the atomic energy levels
\cite{Power1966}. Similarly, the van dar Waals interaction occurs due to the
zero-point fluctuations. The vacuum fluctuations produce a non-vanishing
dipolar moment of the atoms \cite{Spruch1978}. Hence a $1/R^6$ dependent
potential appears in the system. Here $R$ is the distance between the two
atoms. In the relativistic limit, the interaction is further modified by the
influence of the retardation effect \cite{Casimir1948}. It is widely known as
Casimir-polder interaction (CPI). The force associated with this interaction
is known as the Casimir Polder force (CPF), for two parallel conducting plates, the force is attractive, and the expression of the force is given by, $\delta F=-\frac{\pi^2}{240 R^4}$.
Several experimental verifications exist for the
CPI, which proves the Casimir physics as a hallmark in the quantum field theory
\cite{Lamoreaux1996,Mohideen1998}. CPI is also used as an essential tool to
analyze local curvature effects in the presence of quantized fields
\cite{DeWitt1975,Kay1979,Ford1988}. In addition, a thermal character is
connected with the vacuum fluctuations of the quantized fields. It is commonly
known as the thermalization theorem, which tells that if a uniformly
accelerated particle detector interacts with the vacuum state of the external
field and spontaneous emission occurs, then the detector behaves as if it is in
a thermal bath \cite{Unruh1976}. Hence the major implementation of Casimir
physics is shown in thermal-nonthermal scaling of a linearly accelerating atom,
interacting with a massless scalar field
\cite{Marino:2014rfa,Rizzuto:2016ijj,Tian:2014hwa,singha_remarks_2019}. One
major drawback of linear accelerating detectors is that the value of acceleration is
very large to produce 1K temperature \cite{Unruh:1998gq}. In contrast, Bell and
Leinaas identified that in a circular storage ring, one could produce a
relatively higher effective temperature within a shorter time-scale
\cite{BELL1983}. Unlike the uniformly accelerated detector, the rotating
detector cannot achieve a thermal state, as for the circular motion, there is
no existence of event horizon \cite{Letaw1980,Kim1987}. However, due to the
experimental efficiency of the moving electrons in the circular storage ring,
exploring Casimir-physics in the rotating co-ordinates is one of the developing
area in last few decades \cite{Hu2014,HU2015,Cai:2018uck}.\\
Here we consider an atomic system weakly interact with the external quantum fields. These
systems are frequently known as open-quantum system \cite{breuer,atom-photon}.
The reduced dynamics of the system sharply depend on the field correlation
function. There exist several parallel approaches to calculate the atomic
energy level shift due to vacuum fluctuations \cite{Garbrecht_2006}.
Dalibard-Dupont-Cohen (DDC) calculated the rate of change of the mean atomic
energy, which basically determines spontaneous excitation rates and radiative
reactions under the vacuum fluctuations \cite{Dalibard1982}. Similar methods
based on Langevin dynamics are used to calculate the fluctuation induced
interactions \cite{Dean2014}. The quantum master equation (QME) is also an important
tool to derive the reduced dynamical equation of the atoms
\cite{lindblad_generators_1976,kossakowski_quantum_1972}. Benatti first
described the positive time-evolution of the quantum systems weakly coupled
with a scalar field in the Minkowski vacuum \cite{PhysRevA.70.012112}. Later
multiple attempts by using the QME formalism are found to analyze the Casimir
physics from scratch \cite{Tian:2014hwa,Chatterjee_2020,singha_remarks_2019}.
The system-field weak coupling Hamiltonian gives an effective second-order
contribution in the QME. It consists of two parts, the real part gives a
dissipative  dynamics, and the imaginary terms give a second-order shift
term of atomic levels. This shift is known as Lamb shift \cite{breuer}. If the
atoms are interacting with the same external field, then there is a possibility
of the creation of field-induced entanglement, which is known as the
common-environment effect \cite{benatti_environment_2003,Braun2002}. The
individual system-field interaction cross-terms in the second order of QME also
produce a shift term in the dynamics which are particularly dependent on the
distance between the two atoms. As a result, there is a formation of
inter-atomic correlation in the dynamics. This interaction is called resonance
Casimir Polder interaction (RCPI) \cite{Tian:2014hwa}. One of the atoms are
kept in a ground state and the other in an excited state. Due to the
interaction with the vacuum of the same quantized field, the exchange of real
photons occurs, and we are getting a non-zero expectation value of the
correlated state. This phenomena gives rise to resonance interaction
\cite{Rizzuto:2016ijj}.\\ 
In this paper, we briefly discuss the RCPI in the
circular storage ring. We assume the atoms are rotating syn-chronically with
their separation perpendicular to the plane of motion. For a two-atom
system coupled with a massless scalar field, the result is quite
straight-forward \cite{Cai:2018uck}. If we impose some interactions in the
external field, then several non-thermal features come to the scenario. For a
massive scalar field in Schwarzschild spacetime, the RCPI  shows a retarded
behavior\cite{Chatterjee_2020}. In the presence of a vector field (i.e.
electromagnetic field) the response also shows a non-Planck spectrum
\cite{Boyer1980,Hu2014,Rizzuto:2016ijj,She2019}. We simultaneously consider
these two above interactions in our calculations. The energy shifts due to RCPI
are calculated. The shift terms show a unique retarded behavior in comparison
with the massless case. We also derive the length-dependence of the RCPI and
compare it with the thermal behaviors. The energy shift can be encoded by the
polarization transfer technique. For two stationary qubit, quantum information
can be transferred between different polarization modes within the coherence
time \cite{northup_quantum_2014}. That information transfer can be envisaged
using any kind of spectroscopy \cite{Adams1708,PhysRevA.74.042503}. Still, one cannot get ride of the decoherence
effect due to system-environment coupling. However, in the intermediate time
regime or before the system equilibrate, the effectiveness of this procedure
brings a new era in the field of quantum information processing and quantum
technologies \cite{Nagali2009,Leijnse2011}. We also give here a theoretical
analysis to distinguish the retardation effect of RCPI using the concepts of
polarization transfer.

\section{Dynamics of a two-atom system} \label{dyn}

We assume here, two atoms are rotating syn-chronically in a circular orbit and
the perpendicular distance in the rotating plane is defined as $L$, which is fixed \cite{HU2015}. 
The angular speed is $\lambda_0$, $R$ is the radius of the circular orbit. The tangential velocity
is $\mathcal{V}=\lambda_0 R$. Now, we focus on the dynamics in the ultra-relativistic regime
because the high acceleration needed for experimental realization, it can only be achieved in that regime \cite{Kim1987}.
The positions of the atoms in terms of proper time, are written by,
\begin{widetext}
\begin{eqnarray}
&&t_1(\tau) = \gamma \tau,\quad x_1(\tau)= R \cos \frac{\gamma \tau \mathcal{V} }{R},
\quad y_1(\tau)= R \sin \frac{\gamma \tau \mathcal{V} }{R}, \quad z_1(\tau)=0 \nonumber\\ 
&&t_2(\tau) = \gamma \tau,\quad x_2(\tau)= R \cos \frac{\gamma \tau \mathcal{V} }{R}
,\quad y_2(\tau)= R \sin \frac{\gamma \tau \mathcal{V} }{R}, \quad z_2(\tau)=L ~. 
\label{eq:1}
\end{eqnarray}
\end{widetext}
Here $\gamma=1/\sqrt{1-\mathcal{V}^2}$ is the Lorentz factor. The centripetal acceleration is given by
$a=\mathcal{V}^2\gamma^2/R$, which will provide a length scale in the dynamics. The atoms are
weakly coupled with the external quantum fields. The distance between the atoms is taken to be
smaller than the correlation length of the field. Total Hamiltonian of the system+field is written
as \cite{PhysRevA.70.012112},
\begin{eqnarray}
\mathcal{H}= \mathcal{H}_s+\mathcal{H}_f+\mathcal{H}_{sf}~.\nonumber
\end{eqnarray}
$\mathcal{H}_s$ is the free Hamiltonian of the system, it is expressed as,
$\mathcal{H}_s= \omega_0/2\big( \sigma_z \otimes \mathds{1} + \mathds{1}\otimes \sigma_z\big)$. 
The system is taken to be isotropic. $\mathcal{H}_f$ is the normal ordered field Hamiltonian. 
Following the second quantization technique, it can be expressed as,
\begin{eqnarray}
\mathcal{H}_f=\int \frac{d^3k}{(2\pi)^3} E_k a^{\dagger}(k)a(k)~.
\label{eq:-1}
\end{eqnarray}
$E_k$ is the frequency of the field and $a,a^{\dagger}$ is the creation and annihilation operator. 
Exact expression of the operators are depend on the characteristics of the field. 
The system-field coupling Hamiltonian is expressed as,
\begin{eqnarray}
\mathcal{H}_{sf} = \alpha \sum\limits_{\mu=0}^{3} \big[ \sigma_{\mu }^{(1)} \otimes \phi_{\mu}(x_1) 
+ \sigma_{\mu}^{(2)} \otimes \phi_{\mu} (x_2)\big]~, 
\label{eq:2}
\end{eqnarray}
$\alpha$ is the coupling strength. $\phi$ is the field and $x_{1}, \,x_2$ are the individual trajectories
of two atoms which is defined earlier. We define,
\begin{eqnarray}
\phi_{\mu} (x)=\sum^N_{a=1}\big[\chi^a_{\mu} \phi^-(x)\,+\,(\chi^a_{\mu})^{\dagger} \phi^+(x)  \big]~,
\end{eqnarray}
$\phi^{\mp}(x)$ is the negative and positive field operator of the field and $\chi^a_{\mu}$ are the
corresponding complex coefficients \cite{PhysRevA.70.012112}. The system + field formed a closed system. 
To find a dynamical equation of the system, the starting point is the ``Von Neumann-Liouville" equation 
\cite{breuer}. It is given by,
\begin{eqnarray}
\frac{d \rho(t)}{dt}&=& -i [\mathcal{H}_s+\mathcal{H}_f+\mathcal{H}_{sf}, \rho (t)]~.
\label{eq:3}
\end{eqnarray}
The Eq. (\ref{eq:3}) is further expanded to the second order of the perturbing Hamiltonian $\mathcal{H}_{sf}$
and after taking trace over the field variables the reduced dynamical equation of the two-atom system in proper time ($\tau$) can be
written as,
\begin{eqnarray}
\frac{d \rho_s(\tau)}{d\tau}= -i[\mathcal{H}_s + 
\mathcal{H}_{lamb},\rho_s(\tau)]+\mathcal{L}\big(\rho_s(\tau)\big)~. 
\label{eq:4}
\end{eqnarray}
The above equation is called the Lindblad equation or the quantum master equation 
\cite{lindblad_generators_1976,kossakowski_quantum_1972}. The time evolution operator is a one parameter
semi-group and the completely positivity and trace preservation holds. The initial correlation between
the system and field is ignored, it is called ``Born-approximation" \cite{atom-photon}. 
$\mathcal{H}_{lamb}$ is the second order effective Hamiltonian. It produces the shift term in the dynamics.
$\mathcal{L}$ is called the dissipator, which generates the irreversibility in the reduced dynamics of the system. 
The exact mathematical expression of this two quantity is written by,
\begin{eqnarray}
\mathcal{L}\big(\rho_s\big) &=& \sum \limits_{a,b=1}^{2} \sum \limits_{j,k=1}^3 
\gamma^{ab}_{jk}\Big(\sigma _b ^k\rho_s \sigma _a ^j  
-\frac{1}{2}\{ \sigma _a ^j \sigma _b ^k ,\rho_s \} \Big)~,\\
\label{eq:5}
\mathcal{H}_{lamb}&=& -\frac{i}{2} \sum \limits_{a,b=1}^{2} \sum \limits_{j,k=1}^3 
\mathcal{S}_{jk}^{ab}  \sigma _a ^j \sigma _b ^k~,
\label{eq:6}
\end{eqnarray}
Here $\mathcal{S}_{jk}^{ab}$ and $ \gamma^{ab}_{jk}$ is originated from the Fourier transform of 
the two-point correlation function. They are Kramers-Kronig pairs to each-other. 
Due to the common-environment effect, the atoms become entangled through the field correlation function. 
This entanglement has an initial value dependency \cite{PhysRevA.70.012112}.
The field induced shift term is calculated from the Hilbert transforms of the response function of the field.
\begin{eqnarray}
\mathcal{K}^{ab}(\omega_0) = \frac{\mathcal{P}}{\pi i}\int \limits ^{\infty}_{-\infty}d\omega 
\frac{\mathcal{G}^{ab}(\omega)}{\omega-\omega_0}~.
\label{eq:7}
\end{eqnarray}
$\mathcal{P}$ is the Cauchy principal value. $\mathcal{G}^{ab}(\omega)$ is the response function of the field and it is defined as, 
\begin{eqnarray}
 \mathcal{{G}}^{ab}(\omega)&=&\int\limits^{\infty}_{-\infty} d \Delta \tau 
~e^{i\omega \Delta \tau}~{G}^{ab}(\Delta\tau)~.
\label{eq:8}
\end{eqnarray}
Here we assume $\chi^a_{\mu} $ satisfies $\sum\limits^N_{a=1} \, \chi^a_{\mu} \big( \chi^a_{\nu}\big)^{\dagger}\,
=\, \delta _{\mu \nu}.$ So the field correlation functions are diagonal i.e. $G_{ij}^{ab}(x-y)= 
\delta_{ij}G^{ab}(x-y)$, which is given by,
\begin{eqnarray}
 G^{a b}(\Delta \tau)&=&\langle{\Phi}(\tau,\xx_{a}){\Phi}(\tau',\xx_{b})\rangle~,
\label{eq:9}
\end{eqnarray}
Here $\Delta \tau = (\tau-\tau')$. From the K.M.S condition, the detector response function of a linear accelerating 
particle detector in a massless scalar field is given by \cite{Takagi_1986},
\begin{eqnarray}
\mathcal{G}(\omega)= \frac{\vert \omega \vert}{2 \pi} \big{\{ } \theta(\omega) \frac{1}{e^{\omega/T}-1}
+  \theta(-\omega) \big(1+ \frac{1}{e^{\vert\omega \vert/T}-1} \big)\big{ \} }~.
\label{eq:40}
\end{eqnarray}
$T=a/2\pi$, $a$ is the linear acceleration. The first term determine the absorption rate and the second term 
determine the induced and spontaneous emission rate. The Planck factor in the Eq. (\ref{eq:40}) ensures that the 
particle detector perceive a thermal bath of temperature $T$. For a circularly rotating observer, 
the Planck spectrum is replaced by several anonymous factors which impose non-thermality in the system \cite{Kim1987}. 
In the RCPI, only the spontaneous radiative process occurs hence there is no use of the number densities in calculation 
of the response function  \cite{Hacyan_1990,Tian:2014hwa,singha_remarks_2019}. Here we can write the form of 
$\mathcal{S}^{ ab } _ { jk }$ as,
\begin{eqnarray}
 \mathcal{S}^{ ab } _ { jk }=A^{a b}\delta_{jk}-i B^{a 
b}\epsilon_{jkl}\delta_{3 l}-A^{a b}\delta_{3j}\delta_{3k}~,
\label{eq:10}
\end{eqnarray}
where the terms $A^{a b}$ and $B^{a b}$ are given by,
\begin{eqnarray}
 A^{a b}&=&\frac{\lambda^2}{4}\left[\mathcal{K}^{ab} 
(\omega_0)+\mathcal{K}^{ab}(-\omega_0)\right],
\label{eq:11}\\
B^{ab}&=&\frac{\lambda^2}{4}\left[\mathcal{K}^{ab} 
(\omega_0)-\mathcal{K}^{ab}(-\omega_0)\right]~.
\label{eq:12}
\end{eqnarray}
The cross terms of the individual system-field Hamiltonian contribute in the off-diagonal elements of 
$\mathcal{H}_{lamb}$  and the expectation values of symmetric ($\vert E \rangle$) and anti-symmetric states
($\vert A \rangle$) are non-zero, 
which tells that the atoms become entangled in the intermediate time-regime due to Lamb shift. 
The expectation values of this two state is given by,
\begin{eqnarray}
&&\delta E_{S_{LS}}=\langle
E|\mathcal{H}_{lamb}|E\rangle\nn\\
&&=-2~i\left[A_2+A_1\right]~,
\nonumber\\
&&\delta E_{A_{LS}}=\langle
A|\mathcal{H}_{lamb}|A\rangle\nn\\
&&=2~i \left[A_2-A_1\right]~.
\label{shift}
\end{eqnarray}
Here, $A^{11}=A^{22}=A_1$, $A^{12}=A^{21}=A_2$. $A_1$ denotes the self term in the interaction and $A_2$ 
corresponds to cross terms. The resonance Casimir Polder force (RCPF) is defined as, $\delta f = - \delta E / \delta L$. 
Hence we consider only the length($L$) dependent terms for the derivation of RCPI \cite{Chatterjee_2020}. As a result, the contribution from $A_1$ 
in Eq. (\ref{shift}) is neglected. Finally, we define, $\delta E_{S_{LS}}=-\delta E_{A_{LS}}=\delta E$.

\section{Two atoms in a massive scalar field} 

Here we consider two atoms moving
in a circular orbit and weakly interacting with the massive scalar field. The
frequency of the field in Eq. (\ref{eq:-1}) is given by, $E_k=\sqrt{k^2+m^2}$.
Two-point correlation function of the massive scalar field is
\cite{birrell_quantum_1982}, 
\begin{eqnarray} 
G(x,x^{\prime})&\equiv& \langle
0|\hat{\Phi}(t,\x) \hat{\Phi}(t',\x')|0\rangle\nn\\
& =& \int \frac{d^4
k}{(2\pi)^{\frac{3}{2}}} \delta(k^2-m^2)e^{-ik(x-x^{\prime})}~,\nn\\
&=&-\frac{m}{4\pi^2}\frac{K_1 \big(m \sqrt{(t-t^{\prime}-i
\epsilon)^2-(\x-\x^{\prime})^2}\big)}{\sqrt{(t-t^{\prime})^2-(\x-\x^{\prime})^2}}~.\nn\\
\label{eq:14} 
\end{eqnarray}
This is called the positive-frequency Wightman
function. \,$i \epsilon$\, is chosen to be small. For a small mass limit, the
expression in Eq. (\ref{eq:14}) reduced to the case under massless scalar field. On the other
hand, in a high mass limit, the correlation function has an exponential decay
factor, so the RCPI has a similarity with the Yukawa potential in that limit.
We are working in the ultra-relativistic limit ($\gamma>>1$) and in this limit,
using the co-ordinates given in Eq. (\ref{eq:-1}) we get the expression for two-point
correlation function as, 
\begin{eqnarray} 
G^{11}(\Delta \tau)&=&G^{22}(\Delta
\tau)\nn\\ &=& \frac{m}{4 \pi^2}\frac{K_1(m \Delta \tau \sqrt{1+a^2 \Delta
\tau^2/12})}{\Delta \tau \sqrt{1+a^2 \Delta \tau^2/12}}~,\nn\\ G^{12}(\Delta
\tau)&=&G^{21}(\Delta \tau)\nn\\ &=& \frac{m}{4 \pi^2}\frac{K_1(m
\sqrt{\Delta \tau^2(1+a^2 \Delta \tau^2/12)-L^2})}{\sqrt{\Delta \tau^2(1+a^2 \Delta
\tau^2/12) -L^2}}~.\nn\\ \label{eq:15} 
\end{eqnarray} 
The spontaneous emission rate for the massless and the massive case generally does not coincide. The
differences lies in the mass gap of the energy spectra which is independent of
the particle trajectory \cite{PhysRevD.94.104055}. In case of circular motion
in massless scalar field the expressions for response function was calculated
in \cite{Cai:2018uck,HU2015}.  Following the same analogy, the response
function for the massive scalar field is given by, 
\begin{widetext}
\begin{eqnarray}
\mathcal{G}^{11}(E_k)&=&\mathcal{G}^{22}(E_k)\nn\\
&=&\frac{a}{8\sqrt{3} \pi}e^{-2\,\sqrt{3}\,
\frac{\Omega(E_k,m)}{a}}+\frac{\Omega(E_k,m)}{2 \pi};
\qquad(E_k>m)~,\nn\\
\nn\\
\mathcal{G}^{12}(E_k)&=&\mathcal{G}^{21}(E_k)\nn\\
&=& \frac{a}{4 \pi}
\frac{e^{-\frac{\Omega(E_k,m)}{a} \sqrt{6 \sqrt{1+a^2 L^2/3}+6}}}{
\sqrt{6 (\sqrt{1+a^2 L^2/3}+6)(1+a^2L^2/3)}}+ \frac{a}{2\pi
}\frac{\sin\big(\frac{\Omega(E_k,m)}{a}\sqrt{6 \sqrt{1+a^2
L^2/3}-6}\big)}{ \sqrt{6 (\sqrt{1+a^2 L^2/3}-6)(1+a^2L^2/3)}} ;
\qquad(E_k>m)~.  \label{eq:16} 
\end{eqnarray}
\end{widetext}
Here we define $\Omega(E_k,m)=\sqrt{E_k^2-m^2}$. The presence of mass gap is quite
similar with the linear case in massive field \cite{Chatterjee_2020}.
The non-Planck exponential terms in Eq. (\ref{eq:16}) is neglected. The
RCPI for the two atoms in massive scalar field is given by
\cite{Chatterjee_2020}, 
\begin{eqnarray} 
\delta E = \frac{\alpha^2
\mathcal{P}}{2 \pi}\int\limits_m^{\infty} d\,
\Omega(E_k,m)\big(\frac{1}{E_k-\omega_0}+\frac{1}{E_k+\omega_0}\big)\mathcal{G}^{12}(E_k)~.
\label{eq:17} 
\end{eqnarray}
For $m>\omega_0$ the interaction behaves
like Yukawa potential. The interaction becomes short range and decays
beyond a characteristic time scale $1/m$ \cite{Chatterjee_2020}. We are
interested in the other limit, $\omega_0>m$. In this limit the RCPI is
given by, 
\begin{eqnarray} 
\delta E = \alpha^{2}_1
\,a\,\frac{\cos\big(\frac{\sqrt{\omega_0^2-m^2}}{a}\sqrt{6 \sqrt{1+a^2
L^2/3}-6}\big)}{ \sqrt{6 (\sqrt{1+a^2 L^2/3}-6)(1+a^2L^2/3)}}~.
\label{eq:18} 
\end{eqnarray}
All the numerical factors are absorbed in
$\alpha$, so the modified interaction strength is defined as,
$\alpha_1$.

\subsection{The length dependence and retarded response of RCPI} 

The expression in Eq. (\ref{eq:18}) has a dependency on $aL$,
$(aL=\frac{\mathcal{V}^2 L}{(1-\mathcal{V}^2) R})$. We analyze the
resonance interaction in two limiting cases. The condition, $aL<<1$ can
be achieved when the inter-atomic separation is very small with
comparison with the radius of the circular path, $(L<<R)$. Basically
this set up is nearly equal to the linear acceleration case, where
inertial approximation is valid. In this limit, the expression can be
written as, 
\begin{eqnarray} 
\delta E = \alpha_1^2
\frac{\cos\big(\sqrt{\omega_0^2-m^2}L\big)}{L}~.  \label{eq:20} 
\end{eqnarray}
In the small mass limit $(m<<\omega_0)$, the phase lag $(\delta \eta)$
in RCPI with respect to massless case is given by, 
\begin{eqnarray}
\delta \eta = \frac{m^2 L}{2 \omega_0}~.  
\end{eqnarray}
In this limit, the length dependence of RCPI exactly matches with the linear case and
for $m=0$. It matches with the thermal response.  For the other case,
$aL>>1$, which is experimentally easier to achieve in the
ultra-relativistic limits. The RCPI is given by, 
\begin{eqnarray}
\delta E =3^{1/4}\alpha_1^2
\frac{\cos\big(12^{1/4}\sqrt{\omega_0^2-m^2}\sqrt{\frac{L}{a}}\big)}{\sqrt{2}
\sqrt{a L^3}}~.  \label{eq:20.1} 
\end{eqnarray} 
Similarly in the small mass limit $(m<<\omega_0)$, the phase lag w.r.t
massless one is given by, 
\begin{eqnarray} 
\delta \eta =
\frac{12^{1/4 }m^2}{2 \omega_0}\sqrt{\frac{L}{a}}~.
\end{eqnarray}
In this limit the local inertial approximation is violated so non-thermality
arises. The length dependency is also not  equal with the linear acceleration
case \cite{Chatterjee_2020}. The $l/L^2$ dependence is replaced by
$1/\sqrt{aL^3}$.  Here $1/l$ is the linear acceleration. The presence of
centripetal acceleration in the system gives a different result from the linear
case. For $m^2>\omega^2$, in the above equations (\ref{eq:20}, \ref{eq:20.1})
the $\cos\big(\sqrt{\omega_0^2-m^2} \big)$ term is changed by
$\exp\big(-\sqrt{m^2-\omega_0^2}\big)$, so the response for the Yukawa like potential
is also exponentially decaying with the mass  \cite{Chatterjee_2020}.

\section{Two atoms in electromagnetic vector field}

In this section we consider the two atom system is rotating in a circular path and weakly coupled
with the electromagnetic field. The Lagrangian of the EM field is obtained by
\cite{Das2008}, 
\begin{eqnarray}
\mathcal{L}=-\frac{1}{4}F_{uv}F^{uv}-\frac{1}{2}(\partial_v A^v)^2~.
\end{eqnarray} 
Here $ F_{uv}= \partial_u A_v - \partial_v A_u$, $A_u$ is the
electromagnetic vector potential, $A_u=(\phi, \vec{A})$. Here we use the
Feynman gauge \cite{Das2008}. Hence the photon operator can be written as,
\begin{eqnarray} 
A_v(\vec{x})= \int \frac{d^3 p}{(2\pi)^3}\frac{1}{\sqrt{2
\vert \vec{p} \vert}} \sum\limits_{\lambda =0}^3
\epsilon_v^{\lambda}(\vec{p})[a_p^{\lambda} e^{i
\vec{p}.\vec{x}}+a_p^{\lambda \, \dagger} e^{-i \vec{p}.\vec{x}}]~.
\label{eq:19} 
\end{eqnarray} 
Here $\epsilon_v^{\lambda}$ is the polarization
vector. The normalization is defined as,
$\epsilon_v^{\lambda}\epsilon_u^{\lambda^{\prime}}
\eta_{\lambda\lambda^{\prime}} = \eta_{uv}$. The two point correlation function
in the Feynman gauge is written as \cite{Das2008,Hu2014}, 
\begin{eqnarray}
\langle 0 \vert A_i(x)A_j(x^{\prime})\vert 0 \rangle= \frac{\eta_{ij}}{4
\pi^2\{(t-t^{\prime})^2-(\x-\x^{\prime})^2\}}~.  
\end{eqnarray} 
The spins are coupled to the electric field via a dipolar coupling. The dipolar moment of the
individual atoms are given by, $\vec{d}= e \vec{r}$. $e$ is the charge of the
atoms. The dipole moment in terms of Pauli matrices is given by,
\begin{eqnarray} 
\vec{d} = -\sigma^{-} \vec{\mathcal{Y}}^{*}  -\sigma^{+}
\vec{\mathcal{Y}}~.    
\end{eqnarray} 
Here $\vec{\mathcal{Y}}= e \langle E
\vert \vec{r} \vert A \rangle $. The self terms produce a very small shift of
the Zeeman energy levels, so $\langle E \vert \vec{r} \vert E \rangle=\langle A
\vert \vec{r} \vert A \rangle \approx 0$. We only consider the off-diagonal
elements which causes the transition in the atomic energy levels.  The coupling
Hamiltonian is written as \cite{atom-photon}, 
\begin{eqnarray} 
\mathcal{H}_{sf} = \vec{d}_{1}.\vec{E}(x_1)   +  \vec{d}_{2}.\vec{E}(x_2)~.  
\end{eqnarray}
Two spins are identical, so $\vec{d}_{1}=\vec{d}_{2}=\vec{d}_{0}$. Here
$E_i=-\partial A_i/\partial t$. Therefore the two point function of electric
field is given by, 
\begin{eqnarray} 
\langle 0 \vert E_i(x)E_j(x^{\prime})\vert
0 \rangle=(\partial_0 \partial_0^{\prime}\delta_{ij}-\partial_i
\partial_j)\langle 0 \vert A_i(x)A_j(x^{\prime})\vert 0 \rangle~.
\end{eqnarray} 
The two-point correlation function in the ultra-relativistic regime is given by, 
\begin{eqnarray}
G^{11}(\Delta \tau)&=&G^{22}(\Delta \tau)\nn\\ &=&
\frac{1}{\pi^2}\frac{1}{\Delta \tau^4 (1+a^2 \Delta \tau^2/12)^2}~,\nn\\
G^{12}(\Delta \tau)&=&G^{21}(\Delta \tau)\nn\\ &=& \frac{1}{
\pi^2}\frac{1}{(\Delta \tau^2(1+a^2 \Delta \tau^2/12) -L^2)^2}~.
\label{eq:25} \end{eqnarray} 
The Fourier transform of the above Eq. (\ref{eq:25})
is given by, 
\begin{widetext}
\begin{eqnarray}
\mathcal{G}^{11}(E_k)&=&\mathcal{G}^{22}(E_k)\nn\\ 
&=&\frac{2 a^2 E_k}{2 \pi}+\frac{2 E_k^3}{\pi}~,\nn\\
\mathcal{G}^{12}(E_k)&=&\mathcal{G}^{21}(E_k)\nn\\
&=& \frac{a^2}{3\pi}\frac{E_k \cos\big(\frac{E_k }{a} \sqrt{6
\sqrt{1+a^2L^2/3}-6}\big)}{(\sqrt{1+a^2 L^2/3}-1)(1+a^2
L^2/3)}+\frac{4a^3}{6^{3/2} \pi}\frac{\sin\big(\frac{E_k }{a} \sqrt{6
\sqrt{1+a^2L^2/3}-6}\big)}{(\sqrt{1+a^2 L^2/3}-1)^{\frac{3}{2}}(1+a^2
L^2/3)}\nn\\ &&+\frac{8 a^3}{3 \sqrt{6} \pi}\frac{\sin\big(\frac{E_k
}{a}\sqrt{6 \sqrt{1+a^2L^2/3}-6}\big)}{\sqrt{(\sqrt{1+a^2 L^2/3}-1)}(1+a^2
L^2/3)^{3/2}}~.  \label{eq:26} 
\end{eqnarray} 
The contour is chosen in the
upper half of the complex plane. The contribution from the imaginary poles are
neglected as they give an exponential decay. In the Eq. (\ref{eq:17}) if we use
the form of $\mathcal{G}^{12}(E_k)$ from Eq. (\ref{eq:26}), the RCPI of the atom
in the electromagnetic field is then given by, 
\begin{eqnarray} 
\delta E &=& d_{0^{\prime}}^2 \Big(\frac{a^2}{3}\frac{\omega_0 \sin\big(\frac{\omega_0
}{a} \sqrt{6 \sqrt{1+a^2L^2/3}-6}\big)}{(\sqrt{1+a^2 L^2/3}-1)(1+a^2
L^2/3)}+\frac{4a^3}{6^{3/2} }\frac{\cos\big(\frac{\omega_0 }{a} \sqrt{6
\sqrt{1+a^2L^2/3}-6}\big)}{(\sqrt{1+a^2 L^2/3}-1)^{\frac{3}{2}}(1+a^2
L^2/3)}\nn\\ &&+\frac{8 a^3}{3 \sqrt{6}}\frac{ \cos\big(\frac{\omega_0 }{a}
\sqrt{6 \sqrt{1+a^2L^2/3}-6}\big)}{\sqrt{(\sqrt{1+a^2 L^2/3}-1)}(1+a^2
L^2/3)^{3/2}}\Big)~.  \label{eq:28} 
\end{eqnarray} 
\end{widetext}
All the numerical factors
are absorbed in the dipolar coupling strength. The modified dipolar coupling
constant is defined as, $ d_{0^{\prime}}$.

\subsection{The length dependence and retarded response of RCPI}

Following the same manner, if we analyze the
resonance interaction in the two limits, for $aL<<1$, it can be expressed as,
\begin{eqnarray} 
\delta E &=& d_{0^{\prime}}^2\Big( \frac{2\omega_0
\sin(\omega_0 L)}{L^2}+\frac{4\cos(\omega_0 L)}{L^3}+\frac{8 a^2}{3
}\frac{\cos(\omega_0 L)}{L} \Big)~. \nn\\ \label{eq:29} 
\end{eqnarray}
The length scale dependence is quite ambiguous. It consists of three terms. The terms are
also simultaneously proportional to system energy levels and square of
centripetal acceleration. It clearly denotes non-thermality in the region where
local inertial approximation is valid.  In the opposite limit, $aL>>1$, the
expression is, 
\begin{eqnarray} 
\delta E &=& d_{0^{\prime}}^2\Big(\frac{
\sqrt{3}\,\omega_0 \sin\big(12^{1/4}\omega_0
\sqrt{\frac{L}{a}}\big)}{aL^3}\nn\\
 &&+ \frac{\cos\big(12^{1/4}\omega_0
\sqrt{\frac{L}{a}}\big)}{\sqrt{aL^7}}\big(\frac{25\times 3^{1/4}}{3\sqrt{2}
} \big)\Big)~.  \label{eq:30} 
\end{eqnarray}
In this limit the RCPI
also shows a strong non-thermality. The expressions for the RCPI in
both the limits can be expressed as, $\delta E = d_{0^{\prime}}^2
\sqrt{\alpha^2+\beta^2}\cos(\gamma - \delta \eta)$, from here we can
easily extract the expression for the length dependence ($\sqrt{\alpha^2+\beta^2}$) and phase lag ($\delta \eta$).
For $aL<<1$, $\gamma$ has the same form like the massless case
($\gamma=\omega_0 L$). The amplitude and phase lag is given by
\begin{eqnarray} 
\sqrt{\alpha^2+\beta^2}&=& \sqrt{\frac{4 \omega_0^2}{
L^4}+\big(\frac{4}{ L^3}+\frac{8a^2}{3
L}\big)^2}~,\nn\\ \delta \eta &=&
\tan^{-1}\big(\frac{\omega_0 L}{2}\big)~,
\end{eqnarray} 
and for $aL>>1$, ($\gamma=12^{1/4}\omega_0
\sqrt{\frac{L}{a}}$). The phase lag and amplitude is given by,
\begin{eqnarray} 
\sqrt{\alpha^2+\beta^2}&=&\sqrt{\frac{3 \omega_0^2}{
a^2L^6}+\frac{1}{aL^7}\big( \frac{25\times
3^{1/4}}{3\sqrt{2} }\big)^2}~,\nn\\ \delta \eta &=&
\tan^{-1}\big(\frac{3\times 12^{1/4}
\omega_0}{25}\sqrt{\frac{L}{a}}\big)~.  \label{eq:30.1} 
\end{eqnarray}
In both the cases, the the resonance interaction exhibit non-thermal
signatures. The amplitude has a crucial dependency on Zeeman frequency
of the atoms. The response also shows a phase lag in the dynamics.

\section{Instantaneous polarization transfer using RCPI} 

In this section, we give a computation protocol to inspect the retardation effect using the
polarization transfer technique. In particular, as an example, we will discuss about the transfer of different magnetization modes in a system of two spin-1/2 magnetic dipoles.
For single spin system, there exist three possible 
magnetization modes namely $\sigma_x,\, \sigma_y,\, \sigma_z$.
Following the same logic, for two spin system the number of possible
magnetization modes is fifteen. The trace preservation is a constraint in the
system. As the atoms has same energy levels, the number of the independent
magnetization modes reduces to nine. The modes are given by, 
\begin{eqnarray}
\mathcal{F}_i&=& \tr_s \{ (\sigma_i \otimes \mathds{1}+ \mathds{1}
\otimes \sigma_i ) \rho_s \}~,\nn\\ \mathcal{F}_{ii}&=& \tr_s \{(
\sigma_i \otimes \sigma_i ) \rho_s\}~,\nn\\ \mathcal{F}_{ij}&=& \tr_s
\{ (\sigma_i \otimes \sigma_{j}+ \sigma_j \otimes \sigma_i ) \rho_s \}~.
\end{eqnarray} 
Here $i=x,y,z$. We consider here only those elements of the RCPI
Hamiltonian which is responsible for the creation of the inter-atomic
correlation. The form of the Hamiltonian in the Zeeman basis is given by,
\begin{eqnarray} 
\mathcal{H}_{ia}= \frac{\delta E}{2} (\sigma_x \otimes
\sigma_x + \sigma_y \otimes \sigma_y )~.  
\end{eqnarray} 
$\mathcal{H}_{ia}$ is the effective Hamiltonian of the resonance interaction. In the intermediate
time scale, the effect of dissipation can be ignored. Hence the dynamics is
pure unitary. So the magnetization modes are changing with time. The dynamical
equation in interaction frame is written as, 
\begin{eqnarray} 
\frac{d \rho^I_s}{d \tau} = -i[\mathcal{H}_{ia}^I, \rho_s^I]~.  
\end{eqnarray}  
$\rho^I_s$ is the density matrices in interaction frame. For simplicity we neglect the $I$ in the superscript of the density matrices. The RCPI Hamiltonian is unchanged in the interaction frame. In terms of magnetization modes, it is given as, 
\begin{eqnarray} 
\frac{d}{d \tau} (\mathcal{F}_x-\mathcal{F}_y) &=&
\frac{\delta E}{2} (\mathcal{F}_{xz}+\mathcal{F}_{yz})~,\nn\\
\frac{d}{d \tau} (\mathcal{F}_{xz}+\mathcal{F}_{yz}) &=& -\frac{\delta E}{2}
(\mathcal{F}_x-\mathcal{F}_y)~.  \label{eq:35} 
\end{eqnarray} 
Other modes are unchanged under the interaction. For initial magnetization,
$\mathcal{F}_x(0)-\mathcal{F}_y(0)= N_1$ and
$\mathcal{F}_{xz}(0)+\mathcal{F}_{yz}(0)=0$,  the solutions of Eq. (\ref{eq:35})
are expressed as, 
\begin{eqnarray} 
\mathcal{F}_x-\mathcal{F}_y &=& N_1
\cos\frac{\delta E \tau}{2}~, \nn\\ \mathcal{F}_{xz}+\mathcal{F}_{yz} &=& N_1
\sin\frac{\delta E \tau}{2}~.  
\end{eqnarray} 
We can distinguish the energy shift due to 
RCPI for massive and EM field case with the massless case using the
polarization transfer technique. The magnetization transfer time is different
for the interaction of the atoms with various quantum fields. For $\tau_{in}=
\frac{\pi}{\delta E}$, the magnetization switches to
$\mathcal{F}_{xz}(\tau)+\mathcal{F}_{yz}(\tau)$. Here $\tau_{in}$ is the instantaneous
time-scale. We assume $\tau_c<<\tau_{in}<<T_1$. $\tau_c$ is the field-correlation
time, and $T_1$ is the system relaxation time in proper frame. In lab frame the expression for the instantaneous coherence time is expressed as,
\begin{eqnarray}
t_{in}= \frac{\pi}{ \delta E \, \sqrt{1-\lambda_0^2 R^2}}
\end{eqnarray}
 An initial magnetization mode
$\mathcal{F}_x(0)-\mathcal{F}_y(0)$ can be created by using different pulse
sequence which is routinely used in nuclear magnetic resonance spectroscopy
\cite{Furman_2005,BODENHAUSEN1984370}. Similarly, the frequency spectrum in the
Fourier domain gives a clear picture about the different types of responses for
a fixed $L$. The Fourier transform of the temporal response of RCPI is delta
function $[f(\omega)= \delta(\omega - \delta E)]$. So, the various peaks
correspond to different $\delta E$. In the presence of relaxation, the delta
function modifies by a Lorentzian distribution.

\section{Discussions}

As a summary, in this paper, we explore the RCPI of a syn-chronically rotating
two-atom system in the circular storage ring. The atoms move through a
quantized field. They are kept in a ground state and excited state and become
entangled due to the interaction with the common external field, which is
called the resonance interaction. The quantum master equation is the essential
tool to calculate the energy shift due to the resonance interaction.  In the
presence of the massless free scalar field, the thermalization theorem holds
for a uniformly accelerating particle detector. The thermal nature can be
observed when the inertial approximation is maintained. The non-thermal
characters arise when centripetal acceleration present in the system. We can
get back the thermal limit when the radius of the circle is much larger
than the inter-atomic distance. For interaction with the massive scalar field,
the length scale dependency is similar to the massless case. For $aL<<1$, it
has a $1/L$ dependence and for $aL>>1$, it has a  $1/\sqrt{aL^3}$ dependence.
The major aspect of the massive scalar field case is the retarded response. It
is expected that RCPI shows a periodic response. In the presence of mass, we
get a retarded periodic response, which is basically governed by a $m^2$
factor. When the inertial approximation is not valid, then the phase-lag is
also modified by the centripetal acceleration. If we consider an
electromagnetic-vector field, the response function is also corrected by a
$a^2$ term, and the non-Planck factor is also present. As a result, the RCPI
exhibits a non-thermal behavior, which is also independent of the inertial
approximation. In both the limits ($aL>>1, aL<<1$), the amplitude of RCPI
decays much faster than the massless case. The amplitude also depends on the
Zeeman energy of the atoms. A retarded response was also noticed in this case.
The oscillatory behavior of RCPI changes the nature of forces in each period of
the inter-atomic length scale. If we consider this kind of interaction, the
phase lag is present, which can alter the characteristics of RCPI \emph{w.r.t}
the mass-less case. It seems that an attractive RCPF for the mass-less case may
behave like a repulsive force for massive or EM field case and vice versa.
Surprisingly the effect of Zeeman frequency on the phase lag for massive and EM
field cases is different. Increasing the Zeeman frequency, the phase lag for a
massive case goes to zero, whereas it has a constant phase lag ($\delta
\eta=\pi/2$) for the EM field. The characteristics of the interaction also can
be verified by using the polarization-transfer technique. The expectation value
of energy is different for mass-less, massive, and EM field case. The
magnetization transfer time should be different for them, as $t_{in}=\pi/( \gamma\delta
E)$. Similarly, in the frequency domain, the energy peaks appear at different
points, which is also used as an important tool to distinguish the different
responses. Future experimental protocols can be designed by using NMR
spectroscopy to justify the Casimir effect in a circular storage ring.

\begin{acknowledgments}

SS and AC thank University Grant Commission, India, for supporting their work.

\end{acknowledgments}

\end{document}